\documentclass[conference]{IEEEtran}
\IEEEoverridecommandlockouts
\usepackage{cite}
\usepackage{amsmath,amssymb,amsfonts}
\usepackage{algorithm}
\usepackage{algorithmic}
\usepackage{graphicx}
\usepackage{textcomp}
\usepackage{xcolor}
\usepackage{float}
\usepackage{mdframed}
\usepackage{multirow}
\usepackage{url}
\usepackage{amsmath}

\def\BibTeX{{\rm B\kern-.05em{\sc i\kern-.025em b}\kern-.08em
    T\kern-.1667em\lower.7ex\hbox{E}\kern-.125emX}}
\begin{document}

\title{SyntheticPop: Attacking Speaker Verification Systems With Synthetic VoicePops\\}

\author{Eshaq Jamdar$^{1}$, Amith Kamath Belman$^{2}$% <-this % stops a space}
\\Computer Science Department, College of Science, San Jose State University
\\\textit{\{eshaq.jamdar, amith.kamathbelman\}\texttt{@}sjsu.edu}
}

\maketitle

\begin{abstract}
% Voice Authentication (VA), more commonly referred to as automatic speaker verification (ASV), is a popular form of authentication used primarily by automated phone services, such as banking, as a second form of authenticating users. This form of authentication is rife with attacks, such as replay attacks, impersonation attacks, and the growing threat of deepfake audio replicated from the voice of the genuine user. Many solutions have been presented to defend against these types of attacks.  Defense based on "Voice Pops", aims to identify the unique phoneme pronunciations of a person during enrollment. While such a solution is promising, it fails to show how well the system is resilient against other types of attacks, such as more logical based attacks.%
Voice Authentication (VA), also known as Automatic Speaker Verification (ASV), is a widely adopted authentication method, particularly in automated systems like banking services, where it serves as a secondary layer of user authentication. Despite its popularity, VA systems are vulnerable to various attacks, including replay, impersonation, and the emerging threat of deepfake audio that mimics the voice of legitimate users. To mitigate these risks, several defense mechanisms have been proposed. One such solution, ``Voice Pops", aims to distinguish an individual's unique phoneme pronunciations during the enrollment process. While promising, the effectiveness of VA+VoicePop against a broader range of attacks, particularly logical or adversarial attacks, remains insufficiently explored. We propose a novel attack method, which we refer to as SyntheticPop, designed to target the phoneme recognition capabilities of the VA+VoicePop system. The SyntheticPop attack involves embedding synthetic ‘pop’ noises into spoofed audio samples, significantly degrading the model’s performance. We achieve an attack success rate of over 95\% while poisoning 20\% of the training dataset. Our experiments demonstrate that VA+VoicePop achieves 69\% accuracy under normal conditions, 37\% accuracy when subjected to a baseline label flipping attack, and just 14\% accuracy under our proposed SyntheticPop attack, emphasizing the effectiveness of our method.

%system outlined by Wang et al. \cite{b2}, test how well it performs against unseen data, and how well the system withstands data poisoning attacks. The VoicePop system replicated gets an average of 69\% accuracy with another dataset that includes other types of attacks not covered by Wang et al.'s paper. Furthermore, this study dives into two types of attacks on VoicePop: label flipping and adding syntehtic pop noises on audio labeled spoof. Data poisoning, by label flipping, has an unintended effect on regularization and the prevention of possible over-fitting when using a smaller dataset. Lastly, modifying spoof audio to embed synthetic VoicePop proves to be an effective method that reduces overall model accuracy to under 14\%.
\end{abstract}


\section{Introduction}

Voice Authentication (VA) is a critical component of biometric security. It is primarily deployed in automated telecommunication systems, such as telephone banking, healthcare transcription, and smart home devices, where it acts as a secondary layer of authentication. VA systems face two primary attack vectors: physical and logical attacks \cite{b1}. Physical attacks are typically carried out through replay attacks, where an attacker replays a recording of the victim's voice through a speaker to deceive the system. Logical attacks, on the other hand, involve the use of Text-to-Speech (TTS) or machine-generated voice samples, which may be mistaken for authentic voices. Before the recent advancements of voice synthesis and deepfake audio being more accessible to the general public, many attack defenses were developed with replay attacks in mind, since they are the most common and easiest attack to deploy as evidenced by the usage of  ASVSpoof Datasets from 2017 \cite{b11} and 2019 \cite{b4}. In recent years, the rise of deepfake voice technologies, where synthetic voices are generated to mimic individuals, has introduced an additional and increasingly sophisticated challenge to defending VA systems \cite{b10}.

Liveliness detection in biometric systems for modalities such as fingerprint, iris, and face help mitigate some of the dangers of replay attacks. Such detections allow greater confidence in authenticating users properly. The VoicePop defense mechanism for VA can be viewed as a form of liveliness detection implemented in VA+VoicePop\cite{b2}. The basis for VA+VoicePop can be traced back the usage of phonemes as a way to perform liveliness detection\cite{b4} on VAs. The main feature this defense mechanism aims to exploit is the concept of pop noises being a unique identifier for a user's voice. 
%Pop noises can be divided into two categories: easily and hardly heard. Easily heard phonemes are phonemes that give off a distinct pop noise that can easily be picked up on a microphone. Some examples of easily heard pop-noises are the letters "p", "t", and "k". Naturally, hardly heard are phonemes that do not give off an audible pop-noise and examples of this are the letters "m", "n", and "r". During authentication, the system designed by Mochizuki et al. observes the passphrase given by the user and detects if phonemes are present (easy and hardly heard). If a phoneme is classified as easy, and a pop noise is heard at that letter, then the pop-noise is considered legitimate. However, if a phoneme is classified as hard and a pop-noise is heard, the system will reject this pop-noise. While this is a novel concept, the implementation by Mochizuki et al. is susceptible to replay attacks, because a replay attack (with correct alignment of presence of pop noises) is likely to defeat this method. 

%From the novel findings of Mochizuki et al., Wang et al. expands the concept of phonemes for voice authentication by further adding replay attack detection, along with impersonation detection.

%This paragraph is more suitable in related work. 
 
%This paragraph is more suitable in related work. This should be at the end of literature survey.

%consitant citing style
We propose SyntheticPop, a novel attack method aimed at VA+VoicePop system. We recreate the VA+VoicePop system, and investigate how the system holds up against a baseline label flipping attack and our SyntheticPop attack.

\subsection{Key Contributions}
The key contributions of our research work can be summarized as follows:
%rewrite some part of this confirm popnoise is voicepop. 
\begin{itemize}
    \item Establishing an attack mechanism, SyntheticPop, against VA+VoicePop system, leading to a significant decrease in system accuracy
    \item Verified and replicated the results of VA+VoicePop and testing with an large dataset.
    \item Comparative analysis of a baseline label flipping attack against our SyntheticPop attack to emphasize effectiveness. 

\end{itemize}

\section{Related Work}

VA+VoicePop\cite{b2}, is a defense method against replay based spoofing attacks. It is built upon the idea, of using phonemes by using "pop-noises" to verify the claimed identity of the user\cite{b4}. This is done by identifying the unique way in which an individual pronounces certain types of phonemes. For example, when pronounced, the letter 'p' produces a distinct "pop" causing the human tongue and lips to push air out the mouth. The idea is that phonemes, within a given phrase, can be identified within audio. As a result, the intensity of these pop noises can be mapped out and used to authenticate a user. While a replay attack, in theory, has these pop noises present within the audio captured, playing these sounds over a speaker can make it harder for these pops to be recognized by the system. However, solely relying on this detection is not enough, and VA+VoicePop extends this work by using Gammatone Frequency Cepstral Coefficient (GFCC) to scrutinize voice samples further. While some replay attacks can be quickly rejected with the initial VA+VoicePop checks, some voice samples can confuse background noises for pop noises. To counter this issue, GFCC is used to filter down the audio to get the finer details of bursts of air being spoken into a microphone. This, in essence, captures the naturalness of a human speaking into a microphone.

In the VA+VoicePop system, once the passphrase has been correctly identified, the audio segment is then sent to the authentication server. The first part of this process is phoneme segmentation, where phonemes are extracted during the audio clip and where the pops occur in the audio. The audio is transformed via a Short Time Fourier Transform (STFT); this is used to capture the short burst of energy that is expelled at the time the pop noise occurs in the audio. Pop noises happen in the 100 Hz range, and the STFT helps narrow the low frequencies that occur below or equal to 100 Hz. GFCC feature extraction is used for replay attack detection. Pop noises from the previous stage are segmented at the start and end of the clip, with some overlap of background noise. This segment is then sent through a log function to normalize any spikes in audio and smoothens the shape of the energy bursts. A mean of the overall magnitudes is calculated to represent one of the feature vectors to the Support Vector Machines (SVM). In addition, two other features are calculated: the timing changes from pop noise to background ($\delta_1$) and the sudden shift in energy in the recording ($\delta_2$). $\delta_1$  indicates how the transition from pop noise to background (and vice versa) occurs by displaying the magnitude spike in energy being high during the beginning of the pop noise and the magnitude energy decreasing when going back to background noise. $\delta_2$, on the other hand, represents the rate of change that occurs during these two periods. A genuine pop noise will have both of these values at high rates as the voice captured can capture these more significant details because a microphone can pick up these harsh speech transitions naturally. A replay attack, on the other hand, will have much lower values for both deltas as the noise captured by the microphone cannot pick up on these details due to the audio sample being played from a speaker, which causes these details to be silenced or less pronounced. Once the GFCC features are extracted, a SVM classifier is used to identify if the audio sample is from a human or a replay attack.

\begin{figure}[htbp]
\centerline{\includegraphics[width=1.0\linewidth]{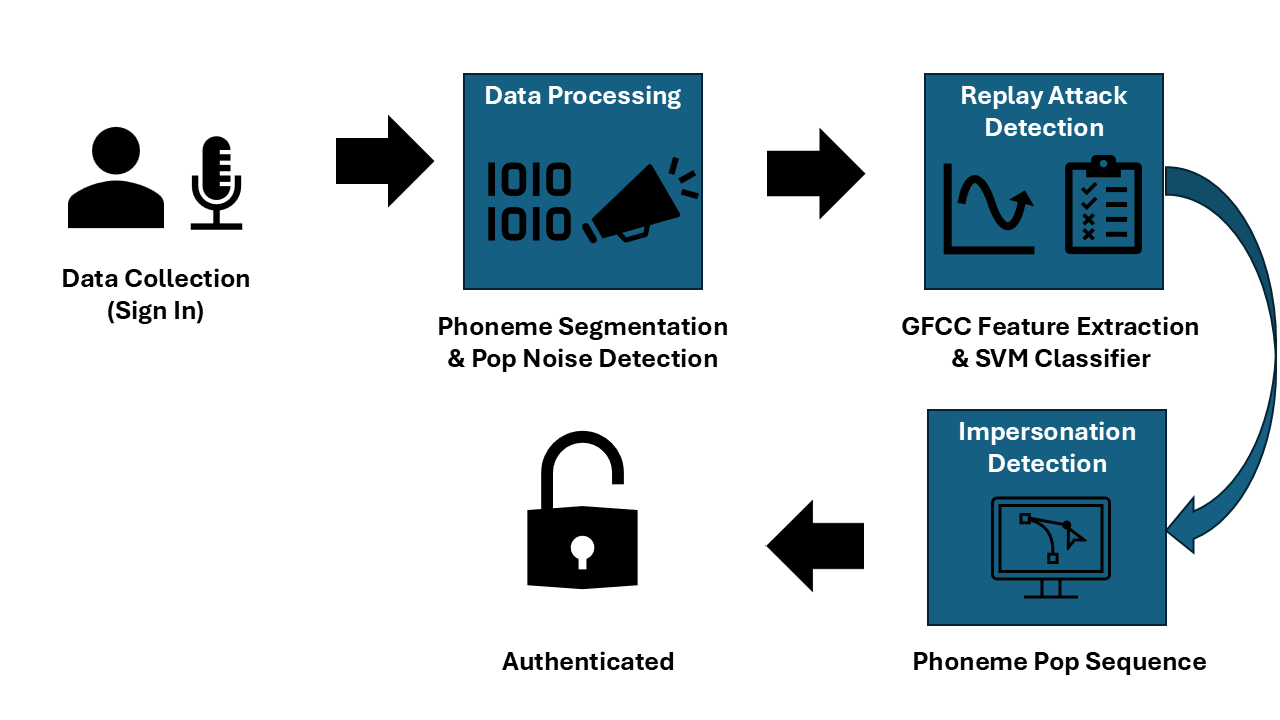}}
\caption{Overview of the VA+VoicePop architecture by Wang et al. \cite{b2}}
\label{fig:1}
\end{figure}

An overview of VA+VoicePop architecture is shown in \ref{fig:1}. While VA+VoicePop work reported a high accuracy of 93.5\% in detecting replay attacks, it has many limitations that require further investigation with logical attacks. In addition, their testing of the system only involved 18 participants, with a majority being men, which brings into question how well the dataset truly is over larger datasets. 

Our work addresses both these concerns, first by recreating VA+VoicePop system and testing it with large external datasets. Second, by deploying our SytheticPop attacks against it to assess robustness.

\section{Methodology}

\subsection{Dataset}
As the dataset used in VA+VoicePop \cite{b2} is not published, we use a larger open source dataset for our experiments. The ASVSpoof 2019 dataset \cite{b4} from AVSSpoof Challenge of 2019 is used as a substitute. A key advantage of using the AVSSpoof dataset is that it consists of the following types of spoof attack audio files:
\begin{itemize}
\item Replay Attacks
\item Deepfake Audio
\item Text to Speech Audio
\end{itemize}
The dataset labels do not differentiate between these types of attacks, as the idea is for the participants to build models that can perform well generally and not for any particular type of spoofing attack. We replicate VA+VoicePop model features for the replay attacks. 

ASVSpoof 2019 contains two labeled sets: train and evaluation. The breakdown for each set can be summed up as:
\begin{itemize}
    \item Train: 2580 Real, 22,800 Spoof
    \item Evaluation: 7355 Real, 63,882 Spoof
\end{itemize}

The publishers intentionally designed both datasets to be imbalanced because this is in line with their philosophy of designing models that perform well against unseen data.

\subsection{Replicating VA+VoicePop}
Since the creators of VA+VoicePop did not provide a repository for any of the code their system ran, this meant having to recreate it. We used the code base from Voice Authentication and Face Recognition \cite{b5} as a starter to build on. This code base provides a notebook environment where users can enroll in their system and authenticate within the system. Mel-frequency Cepstral Coefficients (MFCC) are used to authenticate users.

\begin{algorithm}
\caption{Feature Extraction}
\label{alg:load_voices}
\begin{algorithmic}[1] % The [1] enables line numbering
    \REQUIRE Dataset path \texttt{path}, label dictionary \texttt{labels}
    \ENSURE GFCC features \texttt{gfcc\_stack}, associated labels \texttt{labels}
    
    \STATE Initialize \texttt{gfcc\_features} as an empty list
    \STATE Initialize \texttt{associated\_label} as an empty list
    \STATE Shuffle the items in \texttt{labels} and store in \texttt{shuffled\_labels}
    
    \FOR{\texttt{(fn, label)} in \texttt{shuffled\_labels}}
        \STATE Construct \texttt{voice\_path} as \texttt{path}/\texttt{fn}+".flac"
        \IF{\texttt{voice\_path} exists}
            \STATE Load \texttt{audio} and sample rate \texttt{sr} from \texttt{voice\_path} using \texttt{librosa.load}
            \STATE Extract pop noise \texttt{pop\_noise} from \texttt{audio} and \texttt{sr}
            \STATE Extract GFCC features \texttt{gfcc} from \texttt{audio}, \texttt{sr}, and \texttt{pop\_noise}
            \IF{\texttt{gfcc} is not \texttt{None}}
                \STATE Append \texttt{gfcc} to \texttt{gfcc\_features}
                \STATE Append \texttt{label} to \texttt{associated\_label}
            \ENDIF
        \ENDIF
    \ENDFOR
    
    \IF{\texttt{gfcc\_features} is not empty}
        \STATE Stack \texttt{gfcc\_features} vertically into \texttt{gfcc\_stack}
        \STATE Convert \texttt{associated\_label} to an array
        \RETURN \texttt{gfcc\_stack}, \texttt{labels}
    \ELSE
        \RETURN Empty arrays for both \texttt{gfcc\_stack} and \texttt{labels}
    \ENDIF
\end{algorithmic}
\end{algorithm}

The logic in Algorithm 1 explains the process of Pop Noise detection along with GFCC extraction of features to enable liveliness detection. For every audio file, pop noise timings are extracted, and these timings are then sent to GFCC filters to extract the essential features:
\begin{itemize}
    \item gfcc mean: average of pop noise magnitudes within the audio
    \item $\delta_1$: changes of audio energy from the start and ending of a pop noise
    \item $\delta_2$: how much the pop noise signal changes over the entire signal
\end{itemize}

Once all GFCC features are extracted, the training process can begin. According to VA+VoicePop \cite{b2}, the classifier that uses the GFCC features is an SVM classifier. Since the problem is a binary classification, where audio is either a replay attack (spoof) or real, SVMs are a sound choice. The original VA+VoicePop work used much more balanced dataset that had an equal amount of real and fake samples being trained on their system. However, the dataset is heavily skewed towards spoofs, and to balance the dataset, Synthetic Minority Over-Sampling Technique (SMOTE) \cite{b14} is used to increase the minority class of real audio in the dataset. In addition to increasing the presence of the minority real class, a balanced weight configuration was used to ensure the majority spoof did not overpower the training process. Lastly, a GridSearch was used along with the SVM to identify the proper set of hyper-parameters to use for the best possible model via five-fold cross-validation.

\subsection{Results}
We run two sets of tests: a training of the entire dataset and a training where the maximum number of real (2580) and the same amount but for fakes were used to create a test with a more balanced dataset from the start. Each test was run with the evaluation dataset, which is much larger than the training dataset (7335 real and 63882 fake).

The following metrics are recorded and reported in Table \ref{experiment-table}:
\begin{itemize}
    \item True Positive Rate (TPR): rate of fakes classified as fake
    \item True Negative Rate (TNR): rate of real classified as real
    \item False Positive Rate (FPR): rate of real classified as fake
    \item False Negative Rate (FNR): rate of fake classified as real
\end{itemize}

\begin{table}[h]
\caption{Results of VA+VoicePop With ASVSpoof 2019 Dataset}
\label{experiment-table}
\centering
\begin{tabular}{|c|c|c|c|c|c|}
\hline
\textbf{Experiment} & \textbf{TPR} & \textbf{TNR} & \textbf{FPR} & \textbf{FNR}  &\textbf{Accuracy} \\ \hline
Full Train          &68.69\%          &74.46\%         &25.53\%      &31.31\%   &69.29\%              \\ \hline
Even Train          &39.44\%          &20.86\%         &79.14\%             &60.56\%             &37.52\%                      \\ \hline
\end{tabular}
\end{table}

\begin{figure}[htbp]
\centerline{\includegraphics[width=1.0\linewidth]{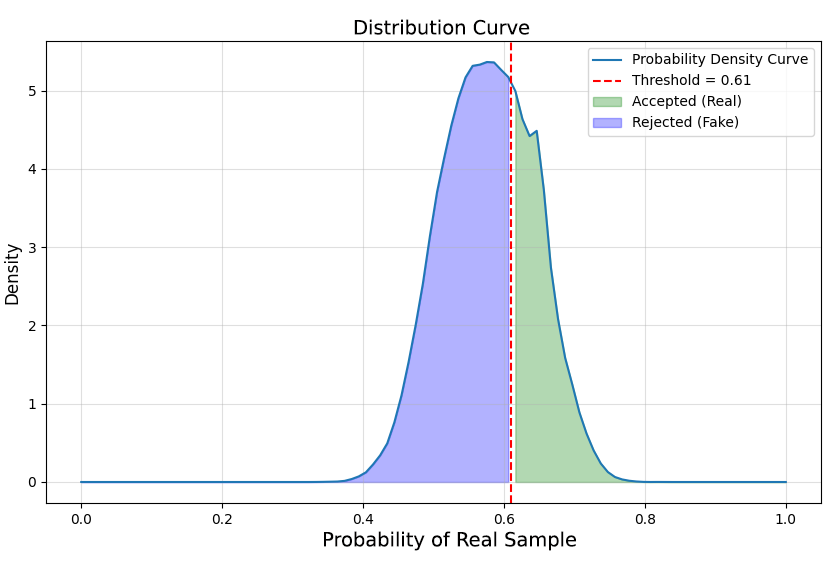}}
\caption{ Probability distribution curve for real samples in Full Training mode}
\label{fig:2}
\end{figure}

\subsubsection{Full Training}
Figure \ref{fig:2} presents the probability distribution of audio being fake or real, using a threshold of 0.61 for the evaluation set aligned the most with the dataset, mostly rejecting and accepting only a small number of the dataset. However, Table~\ref{experiment-table}'s Full Train row shows the results for how well the model aligns with the evaluation dataset. In the context of identifying fakes versus real, the model tends to identify most of the real and most of the fake, although with significant margins of error. 

The model, with full training, has an issue with high FNR. Over 20,000 samples are considered real, and a plausible explanation is that since the dataset includes logical attacks, the VA+VoicePop algorithm focuses exclusively on physical-based attacks. However, this cannot be definitively concluded, as further testing with a dataset solely focused on physical-based attacks is needed to fully confirm this suspicion.
\begin{figure}[htbp]
\centerline{\includegraphics[width=1.0\linewidth]{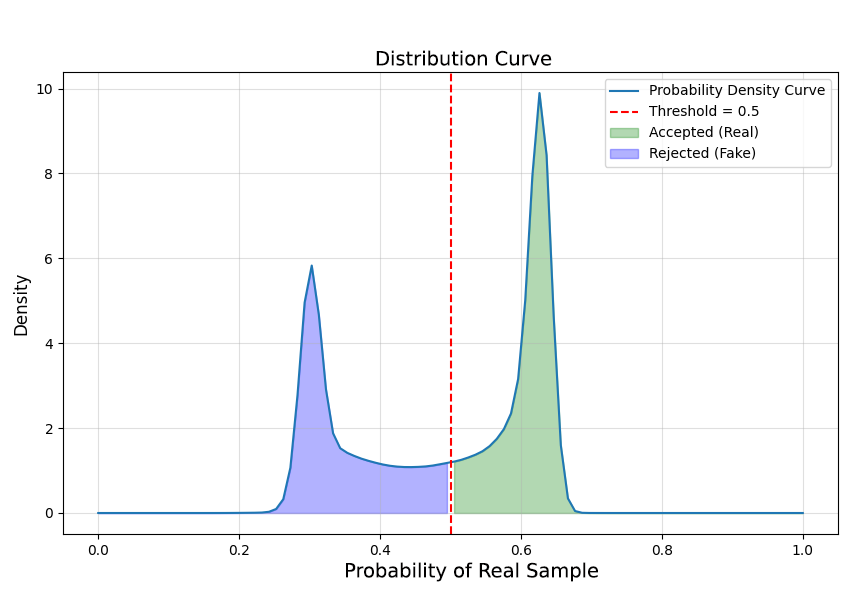}}
\caption{Probability distribution curve for real samples in Even Training mode} 
\label{fig:3}
\end{figure}
\subsubsection{Even Training}
The probability distribution in Figure \ref{fig:3} shows a much more bimodal distribution, where the training caused a clear line of separation between real and fake. The threshold chosen for this test was 0.5 to distinguish between spoof and real.

For the purposes of exploring a more balanced training set, the max number of real samples in the training set, 2580, is set for both the real and fake limiters. This leaves the test with a total of 5160 labels total, unlike the previous 22800 from the whole train set.

Under a balanced dataset with a 0.5 threshold, the accuracy of the model suffers quite a lot in identifying real and fake. From further testing, a higher threshold of 0.6 results in more fakes being classified correctly, but the number of reals being classified correctly continues to decline. Since the number of fake samples to train on is limited by the total number of real ones to test on, and the samples are randomly chosen (meaning we cannot control for a logical or replay-based attack), the model suffers from a form of under-training, or somewhat imbalanced within the confines of the training. A more realistic approach is to stick with the complete training of the model with the entire dataset and fine-tuning to improve the model.

%Small change here. Can be have one showing baseline attack, other SyntheticPop attack. Need to change name in figure. 
\section{Attack Experiment Design}
\begin{figure}[htbp]
\centerline{\includegraphics[width=1.0\linewidth]{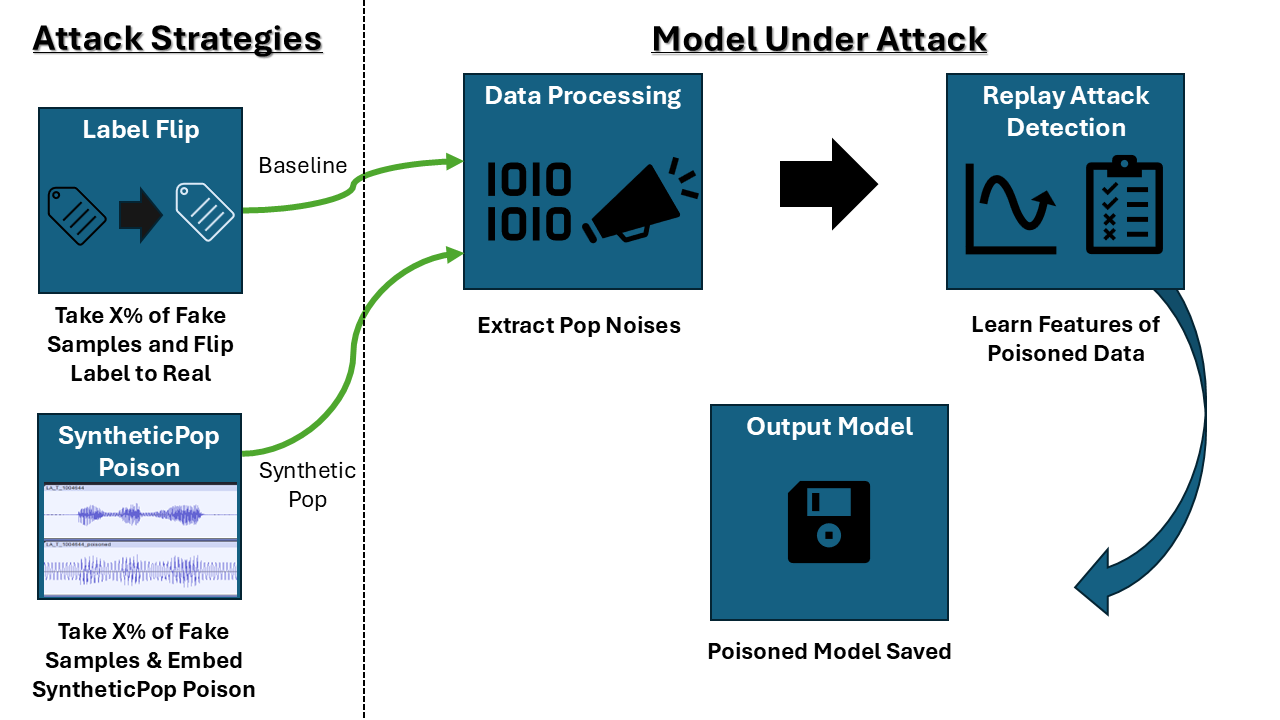}}
\caption{High Level Overview of Attack Strategies}
\label{fig:4}
\end{figure}
Since we have established a baseline for the system accuracy, where on a full training, 74.4\% of real samples are classified as real and 68.6\% of fakes are classified as fakes, this opens up the ability to test how well the system can withstand data poisoning attacks. We chose two approaches, a) a simple label flipping attack to set the baseline for attack success, and b) our novel SyntheticPop attack method. A high level overview of the attack strategies is presented in \ref{fig:4}.

\subsubsection{Label Flipping Attack}
The label flipping attach is a straightforward method that can target a certain amount of labeled data within a population and negate the labels \cite{b7}. These types of attacks are commonly used with Binary Classification problems, making it a good candidate for VA+VoicePop. 

\subsubsection{SyntheticPop Attack}
SVMs can be targeted by altering training data in the hopes that these modifications are enough to disrupt the classification process \cite{b8}. Considering there is an abundance of fake samples during training, a small subset of the population can be modified further by adding synthetic pop noises for a more potent attack. Attack methods where audio samples of a particular user are targeted via perturbations in the audio in specific regions have shown promising results\cite{b12}. We design our white-box assumption based attack algorithm in a similar direction.

%add mathematic equation
\begin{algorithm}
\caption{SyntheticPop poison generation}
\label{alg:modify_pop_freq}
\begin{algorithmic}[1]
\REQUIRE Audio signal ($a$) and sampling rate ($sr$)
\ENSURE Modified audio signal with an amplified pop noise

\STATE Set Amplitude $A$ = 0.5
\STATE Set Frequency $f$ = 90
\STATE Set Audio Length $N$ = len(audio)
\STATE Calculate the Duration $D$ = $\frac{N}{sr}$
\STATE Create Time Vector $t$ as $t$[i] = i $\ast$ $\frac{i * D}{len(a)}$, i = 0,1,2,3
\STATE Generate a sine wave $P(t)$ = $A$ * $\sin$(2$\pi$ * $f$ * $t$) 
\STATE Add $P$ to the audio signal $a$ and generate poison $a_p$ = $\frac{a + P(t)}{\max(\|a + P(t)\|)}$
\RETURN $a_p$
\end{algorithmic}
\end{algorithm}

\begin{figure}[htbp]
    \centering
    \includegraphics[width=0.7\linewidth]{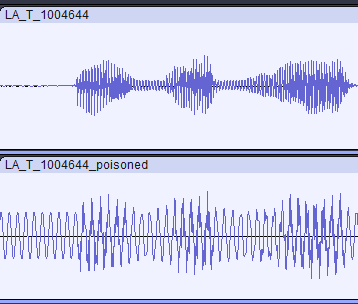}
    \caption{Top: Unpoisoned spoofed audio file, Bottom: The same file after SyntheticPop poisoning}
    \label{fig:5}
\end{figure}

Algorithm 2 describes our SyntheticPop method of generating poisoned spoof samples. As pop noise is a sudden burst of energy that is below or equal to 100 Hz on lower bands, a subtle sine wave can be added to resemble pop noise in a fake audio sample. The control variables in the algorithm dictate the amplitude and frequency values of the signal. They can be fine tuned to be picked up by the VA+VoicePop and GFCC extraction to take place. There is an assumption that there are some pop noises in fake audio samples that occur even in fake audio clips, so the frequency and amplitude are at moderate levels throughout the entire audio clip. 

As seen in Figure \ref{fig:5}, the bottom audio graph is the result of the algorithm applied broadly over the entire audio clip. By  targeting the entire fake audio file with SyntheticPop, the goal is to have the system accept the fake files in the Data Processing Stage and then extract features of the fake audio files. The ultimate effect would be the system being confused on how to classify real or fake sample, thus leading to reduced system accuracy. 

\section {Attack Experiment Results}

Two attacks were deployed on training data, 20\%  poisoning with label flipping and 20\% poisoning using SyntheticPop. 

\subsection{Label Flipping Attacks} 
Each test run randomly selects a sample's label and flip it for the run. For simplicity's sake and due to the long training of the entire dataset, the setup for this test is the maximum of 2580 real from the dataset and 2580 fakes that are randomly obtained due to the abundance of fake samples in the training set and for quicker testing. For the sake of fair comparison, Figures 3 \& Table~\ref{experiment-table}'s even train row will be the standard baseline for comparison, as all poisoning tests use an even number of fake and real samples. 

\begin{table}[h]
\caption{Results of Poisoning VA/VoicePop}
\label{experiment-table-2}
\centering
\begin{tabular}{|c|c|c|c|c|c|}
\hline
\textbf{Experiment} & \textbf{TPR} & \textbf{TNR} & \textbf{FPR} & \textbf{FNR}  & \textbf{Accuracy}           \\ \hline
Label Flip           &39.04\%          &21.32\%       &78.68\%       &60.96\%     &37.21\%           \\ \hline
SyntheticPop            &4.12\%           &100\%         &0\%          &95.88        &14.01\%     \\ \hline

\end{tabular}
\end{table}

\begin{figure}[htbp]
\centerline{\includegraphics[width=1.0\linewidth]{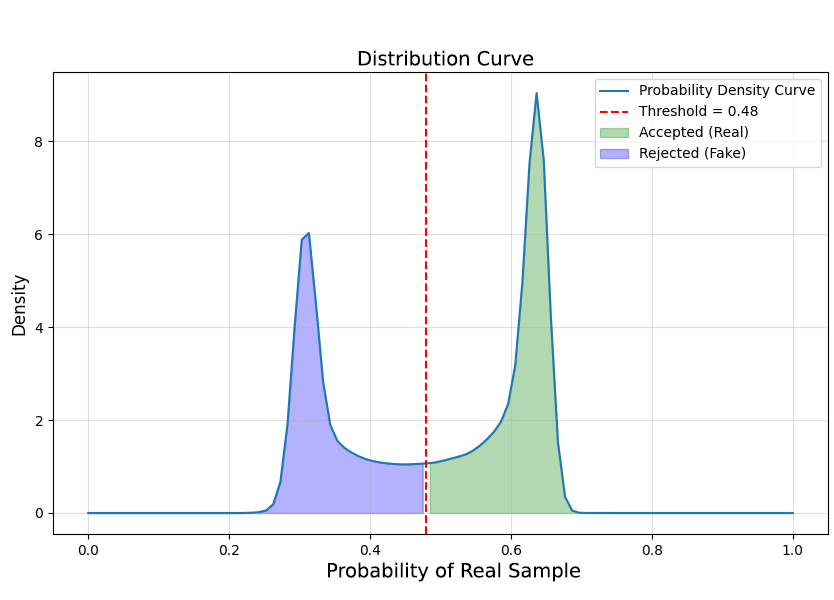}}
\caption{Probability distribution curve for real samples 20\% Poison with Label Flipping}
\label{fig:6}
\end{figure}
Figure \ref{fig:6} shows the overall acceptance threshold for this model (0.48), which is not too far off from Figure \ref{fig:3} because shows a clear line of division for fakes and reals, similar to Figure \ref{fig:3}. It should be noted that the threshold was slightly lowered by -0.2 compared to the 0.5 in \ref{fig:3}. 

Table~\ref{experiment-table-2}'s 20\% poison with label flipping are similar to Table~\ref{experiment-table}'s even train result. Calculating accuracy difference based on Table~\ref{experiment-table}'s even train row, poisoning 20\% of the samples yields an overall accuracy of 37.21\%, a -0.31 difference. Furthermore, the TPR saw a difference of -0.4\%, and FNR increased by 0.4\%. From a preliminary glance, poisoning 20\% of this population gave a slight decrease in overall accuracy as the results still mostly line up with the results of no poisoning. In contrast to fake samples suffering from accurate identification, TNR saw an 0.46\% increase, and the FPR saw a 0.28\% increase over the previous non-poison version. This is in-line with some studies that show, having a higher percentage of flipped labels will have the unintended effect of regularizing \cite{b9}.

It would seem that the label flipping at 20\% is subtle, and even with the increases in identifying accurate samples, the difference is small. The Attack Success Rate (ASR) shows a 60.96\% success rate. However, considering this attack did not deviate significantly from the even training with no poison, the attack statistic can be discounted. 

\subsection{SyntheticPop Poisoning}
%can legend have increased font. in all density curves
\begin{figure}[htbp]
\centerline{\includegraphics[width=1.0\linewidth]{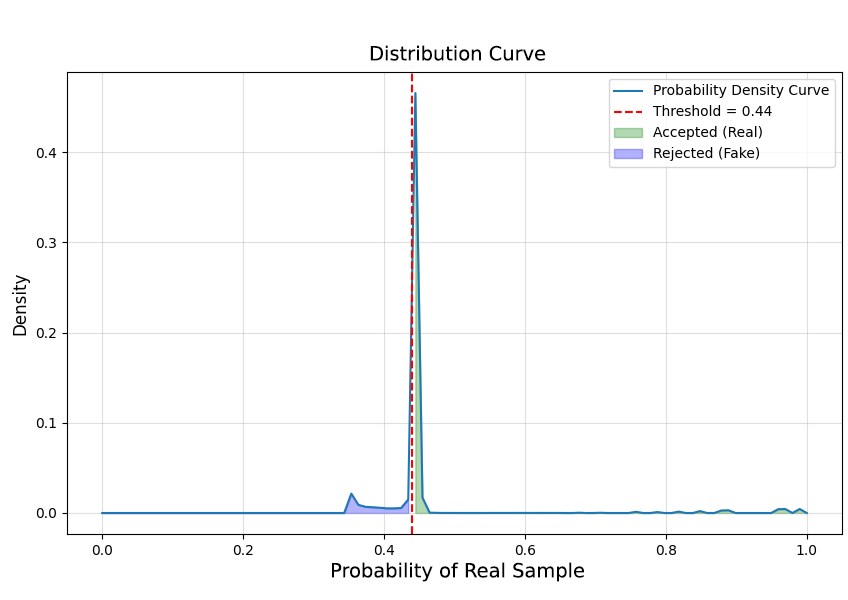}}
\caption{Acceptance Rate of 20\% Poison with SyntheticPop}
\label{fig:7}
\end{figure}
For the purposes of this attack, 20\% of audio samples labeled spoof were randomly selected and modified according to Algorithm 2. 
\subsubsection{Results}
Figure \ref{fig:7}, shows the distribution of probabilities is mostly centered around 0.44 and shows a strong bias towards real samples being accepted over fake. The behavior exhibited here is intriguing, and Table~\ref{experiment-table-2} provides better insight into how well this attack worked. 

Considering the accuracy in Table~\ref{experiment-table-2} in comparison to the label flip attack statistics and even training from Table~\ref{experiment-table} (as both results are very similar to one another), the model's accuracy has fallen to an accuracy 14.01\%, with a TPR decrease of over 35.32\% and the FNR increased by the same amount. For total real samples, all of the real samples were classified as real, with a 79.14\% increase in TNR.

 SyntheticPop attacks caused the model to classify many fakes as real. The number of fakes being classified as accurate is almost a majority of faked labeled samples, and only a smaller number of fakes were correctly identified as fake. Considering that this was only 20\% of the dataset being poisoned if a larger poison population was utilized, it is likely that every audio sample would be treated as accurate.

In terms of the ASR, 95\% of all attack samples made it through, indicating a very high success rate. A revisit to Figure \ref{fig:5} provides better clarity. Before the poisoning, there was not much of a pop noise presence, and there was not much of a pattern that could be picked up by the VA+VoicePop system. However, with the bottom graph, we see that our synthetic pop noises can be observed throughout the entire audio file. The idea is that if you make a noticeable pattern that can be picked up as pop noise during extraction, this would be enough to introduce a noise that will confuse the model and ultimately cause the results seen in Figure \ref{fig:7} and Table~\ref{experiment-table-2}.

\section{Conclusion}

 Key insights from our research can be summarized as follows: a) Our proposed SyntheticPop attack on VA+VoicePop exposed a new vulnerability in the system. VA+VoicePop showed resilience to simple label flipping attacks but was susceptible to SyntheticPop. Under label flipping attacks, fake samples with flipped labels do not cause a large difference in VA+VoicePop accuracy. b) SyptheticPop attacks deployed directly, cause the model to degrade in classification and ultimately resulted in a drastic reduction in VA+VoicePop  accuracy. This type of attack proved to be a determent to the overall system and could have possibly allowed other types of non-replay attacks to be classified as accurate due to this noisy data. The result of such an attack provides insight into the robustness of, or the lack thereof, VA+VoicePop to data poisoning attacks. This is an important finding, as many Voice Recognition systems that can differentiate between human, bots, or replay attacks can be vulnerable to these issues depending on where they receive such data.

The use of ASVSpoof 2019 dataset presents subtle limitations to our work. Further studies with datasets focused on single type of attacks, like either replay or deepfake, but not both, would help found a clearer understanding of these vulnerabilities in VA systems. Attack testing incorporating diffusion based deepfake audio with pop noises would help analyze the effect of cloning samples on a VA+VoicePop system. Although some studies show generative models can sometimes make audio sound too unnatural and too clean \cite{b13}. It would be intriguing to implement approaches like SyntheticPop described in this study to explore the transferability of such attacks to different modalities and biometrics.


\begin{thebibliography}{00}

\bibitem{b1} J. Zhou, T. Hai, D. N. A. Jawawi, D. Wang, E. Ibeke, and C. Biamba, "Voice spoofing countermeasure for voice replay attacks using deep learning," \emph{Journal of Cloud Computing: Advances, Systems, and Applications}, vol. 11, no. 51, pp. 1--14, 2022. [Online]. Available: \url{https://doi.org/10.1186/s13677-022-00306-5}

\bibitem{b2} Q. Wang, X. Lin, M. Zhou, Y. Chen, C. Wang, Q. Li, and X. Luo, "VoicePop: A pop noise-based anti-spoofing system for voice authentication on smartphones," in \textit{Proceedings of the IEEE International Conference on Communications (ICC)}, 2024. [Online]. Available: \url{https://doi.org/10.48550/arXiv.2007.08199}.


\bibitem{b3} X. Valero and F. Alías, "Gammatone cepstral coefficients: Biologically inspired features for non-speech audio classification," \emph{IEEE Transactions on Multimedia}, vol. 14, no. 6, pp. 1684--1692, Dec. 2012. [Online]. Available: \url{https://doi.org/10.1109/TMM.2012.2199972}

\bibitem{b4} J. Yamagishi, M. Todisco, M. Sahidullah, H. Delgado, X. Wang, N. Evans, T. Kinnunen, K. A. Lee, V. Vestman, and A. Nautsch, "ASVspoof 2019: The 3rd Automatic Speaker Verification Spoofing and Countermeasures Challenge database," \emph{University of Edinburgh, The Centre for Speech Technology Research (CSTR)}, 2019. [Online]. Available: \url{https://doi.org/10.7488/ds/2555}

\bibitem{b5} S. Mochizuki, S. Shiota, and H. Kiya, "Voice liveness detection based on pop-noise detector with phoneme information for speaker verification," \emph{Journal of the Acoustical Society of America}, vol. 140, no. 4, pp. 3060, 2016. [Online]. Available: \url{https://doi.org/10.1121/1.4969520}

\bibitem{b6} M. Merchant, \emph{Voice Authentication and Face Recognition}, GitHub repository, 2024. [Online]. Available: \url{https://github.com/MohamadMerchant/Voice-Authentication-and-Face-Recognition}.

\bibitem{b7} Q. Xu, Z. Yang, Y. Zhao, X. Cao and Q. Huang, "Rethinking Label Flipping Attack: From Sample Masking to Sample Thresholding," in IEEE Transactions on Pattern Analysis and Machine Intelligence, vol. 45, no. 6, pp. 7668-7685, June 2023. [Online] Available: \url{https://doi.org/10.1109/TPAMI.2022.3220849}

\bibitem{b8} M. A. Ramirez, S.-K. Kim, H. Al Hamadi, E. Damiani, Y.-J. Byon, T.-Y. Kim, C.-S. Cho, and C. Y. Yeun, "Poisoning attacks and defenses on artificial intelligence: A survey," \emph{arXiv preprint arXiv:2202.10276}, Feb. 2022. [Online]. Available: \url{https://arxiv.org/abs/2202.10276}

\bibitem{b9} H. Song, M. Kim, D. Park, J. Shin, and J.-G. Lee, "Learning from noisy labels with deep neural networks: A survey," \emph{ResearchGate}, Mar. 2022. Available: \url{https://doi.org/10.1109/TNNLS.2022.3152527}

\bibitem{b10} A. Mittal and M. Dua, "Automatic speaker verification systems and spoof detection techniques: review and analysis," \emph{International Journal of Speech Technology}, Aug. 2021. [Online]. Available: \url{https://doi.org/10.1007/s10772-021-09876-2}

\bibitem{b11} T. Kinnunen, M. Sahidullah, H. Delgado, M. Todisco, N. Evans, J. Yamagishi, and K. A. Lee, "The ASVspoof 2017 Challenge: Assessing the Limits of Replay Spoofing Attack Detection," \emph{Interspeech 2017}, Aug. 2017. [Online]. Available: \url{https://doi.org/10.21437/Interspeech.2017-1111}

\bibitem{b12} G. Chen, S. Chen, L. Fan, X. Du, Z. Zhao, F. Song, and Y. Liu, "Who is Real Bob? Adversarial Attacks on Speaker Recognition Systems," \emph{arXiv preprint}, Apr. 2020. [Online]. Available: \url{https://doi.org/10.48550/arXiv.1911.01840}

\bibitem{b13} C. Zhang, C. Zhang, S. Zheng, M. Zhang, M. Qamar, S.-H. Bae, and I. S. Kweon, "A Survey on Audio Diffusion Models: Text-To-Speech Synthesis and Enhancement in Generative AI," \emph{arXiv preprint}, Apr. 2023. [Online]. Available: \url{https://doi.org/10.48550/arXiv.2303.13336}

\bibitem {b14} N. V. Chawla, K. W. Bowyer, L. O. Hall, and W. P. Kegelmeyer, 
“SMOTE: synthetic minority over-sampling technique,” 
Journal of Artificial Intelligence Research, vol. 16, no. 1, pp. 321–357, Jun. 2002. [Online]. Available: 
\url{https://doi.org/10.5555/1622407.1622416}



\end{thebibliography}
\end{document}